\documentclass[aps,preprint]{revtex4}
\usepackage[dvipdfm]{graphicx,color}
\usepackage{dcolumn}
\usepackage{bm}
\usepackage{amsmath,amsthm,amssymb}
\usepackage{spintorque}

\begin{document}
\title{Current-induced domain wall motion in Rashba spin-orbit system}
\author{Katsunori Obata$^{1}$}
\author{Gen Tatara$^{1,2}$}
\affiliation{%
$^{1}$Department of Physics, Tokyo Metropolitan University,
Hachioji, Tokyo 192-0397, Japan
\\
$^{2}$PRESTO, JST, 4-1-8 Honcho Kawaguchi, Saitama 332-0012, Japan
}

\date{\today}


\begin{abstract}
Current-induced magnetic domain wall motion,
induced by transfer of spin transfer effect due to exchange interaction, 
is expected to be useful for next generation high-density storages.
We here show that efficient domain wall manipulation can be achieved by introduction of Rashba spin-orbit interaction, which induces spin precession of conduction electron and acts as an effective magnetic field.
Its effect on domain wall motion depends on the wall configuration.
We found that the effect is significant for Bloch wall with the hard axis along the current, since the effective field works as $\beta$ or field-like term and removes the threshold current if in extrinsic pinning is absent.
For N\'eel wall and Bloch wall with easy axis perpendicular to Rashba plane, the effective field induces a step motion of wall corresponding to a rotation of wall plane by the angle of  approximately $\pi$ at current lower than intrinsic threshold.
Rashba interaction would therefore be useful to assist efficient motion of domain walls at low current.
 
\end{abstract}
\maketitle

\section{Introduction}
In recent years, magnetic-memory devices like hard disk drives have been utilized for various products such as portable music players and home appliances.
 With the diversification of the use of the devices, they deal with a much larger amount of information and need to be more miniaturized and have higher capacity.
 However, it is believed that the conventional magnetic-memory devices will reach the limit of downsizing and high capacity in the near future.
 The reason is because the magnetization so far is controlled by a magnetic field and the more miniaturized and the higher capacity they get, the higher necessary field they will need.
 Therefore, a new method which replaces the magnetic field is required.
 One of the methods expected is the control by electric current, i.e. current-induced magnetization reversal.
The current-induced magnetization reversal was pointed out by Berger~\cite{Berger78,Berger86,Berger96} and Slonczeski.~\cite{Slonczewski96}
The torque is caused by the $s$-$d$ exchange interaction via spin transfer effect arising from conservation of spin angular momentum between the current and the magnetization.

To realize devices using the current-induced magnetization reversal, it is absolutely essential to reduce the current necessary.
In this paper, we demonstrate that the problem can be solved by introduction of Rashba spin-orbit interaction.~\cite{Rashba}
A spin precession induced by Rashba interaction is expected to affect the spin transfer torque mechanism, and to lead to efficient control of magnetization.
The interaction has been proposed on two-dimensional electron systems realized at the interface of semiconductors, but is now known to arise quite generally when inversion symmetry is broken.~\cite{Lashel96}
For instance, significant Rashba effect can arise on surface of heavy metals.~\cite{Nakagawa07,Ast07}
Our results therefore would apply to systems of magnetic semiconductors~\cite{yamano} with gate for Rashba interaction attached, or on magnetic metallic thin films with heavy ions doped.~\cite{Ast07}

Recent theoretical studies revealed 
that current-driven domain wall motion is significantly affected by spin relaxation process.
Actually, spin relaxation triggers a torque perpendicular to the adiabatic spin transfer torque~\cite{Zhang04,Thiaville05} and this torque, $\beta$ torque~\cite{Thiaville05},  acts as an effective force on domain wall, which deletes intrinsic pinning effects\cite{Thiaville05,TTKSNF06}.
Microscopic analysis of spin relaxation was done in the case of spin flip scattering by random impurity spins and value of $\beta$ was found to be of similar magnitude with (but not necessarily equal to) the spin relaxation contribution to Gilbert damping parameter $\alpha$ \cite{Tserkovnyak06,KTS06,KS07,Duine07}.
On the other hand, role of spin-orbit interaction, which also causes spin relaxation, on the wall dynamics has not been much studied so far.
Spin-orbit interaction in magnetic semiconductors was recently studied based on  Kohn-Luttinger Hamiltonian, and large enhancement of wall velocity due to spin-orbit interaction was found~\cite{Nguyen07}.
The Rashba spin-orbit interaction we are going to study here turns out also to assist wall motion to a large extent.
In this paper, we study current-driven domain wall motion in the presence of Rashba interaction.
Rashba interaction acts on $x$-$y$ plane and electric current  is applied along $x$ axis.
The spin system we consider is with easy and hard axes.
The wall is treated as planar and rigid, which is justified if easy axis energy gain, $K$, is larger than hard axis anisotropy energy, $K_\perp$\cite{begt2,TKSLK07,TKS08}.
We consider three cases where easy axis direction (we call $\eta$) is $x$, $y$ and $z$. 
The wall structure in these three cases are N\'eel for $\eta=x$ and Bloch wall for $\eta=y, z$. 
We will see that dynamics of N\'eel wall and Bloch wall with $\eta=z$ (called Bloch(z)) are essentially the same, while Bloch with $\eta=y$ (Bloch(y)) is different. 
The effect of current is calculated for each anisotropy configuration using gauge transformation in spin space, assuming adiabatic limit.
The Rashba interaction is treated perturbatively to the second order. 
Expansion with respect to Rashba interaction is justified here since electron has large spin polarization $\Delta$.
(This is in contrast to the spin Hall case without polarization, where perturbative treatment is not allowed~\cite{Inoue}.) 
Below, we derive the effective Hamiltonian for local spin,  derive the equation of motion for domain wall, and then discuss wall dynamics solving the equation.

\section{Model and method}
\subsection{Local spin}
Domain wall is described by a Lagrangian of local spins given by
\eq{
L_s = \inthra \hbar S \lc{\cos{\theta}-1} - H_s \label{eq:ls2},
}
where local spin direction is expressed by polar coordinates ($\theta , \phi$).
The first term describes time evolution of local spin (spin  Berry phase term).
The  Hamiltonian of local spin we consider is given as follows;
\eq{
H_s = \inthra \lc{\frac{J}{2} (\nabla S)^2 -\frac{K}{2} (S_{\|})^2 + \frac{K_{\bot}}{2} (S_{\bot})^2}. 
\label{eq:hs1}
}
The first term is ferromagnetic exchange interaction between local spins, the second and third terms are magnetic anisotropy energies.
$S_{\|}$ is the easy axis component of spin and $K(>0)$ is the corresponding anisotropic energy, and 
$S_{\bot}$ and $K_{\bot}$ are hard axis ones.

In the absence of Rashba interaction, we can choose space coordinate and spin coordinate independently.
When Rashba spin-orbit interaction is switched, spin and space coordinates correlate each other,
and the spin torque and gauge field depend on the choice of magnetic easy axis, which we call $\eta$ axis.
We consider three different cases with easy axis $\eta = x,y$ and $z$. 
Domain wall structure then becomes N\'eel and Bloch walls. 
We will call domain wall configuration with the easy axis label, such as  Bloch(z) for a Bloch wall with $\eta =z$
 (see Table. \ref{dwhard}).
\begin{table}[h]
\centering
\begin{tabular}{lcc} \hline \hline
& \hspace{0.05 \linewidth} Easy ($\eta$) axis  ($K$) \hspace{0.05 \linewidth} & Hard axis ($K_{\bot}$)  \hspace{0.05 \linewidth} \\ \hline
Bloch(z) & z & y  \\ \hline  
Neel(x) & x & z \\ \hline
Bloch(y) & y & x \\ \hline \hline
\end{tabular}
\caption{
Configuration of magnetic anisotropy and corresponding domain wall structure.}
\label{dwhard}
\end{table}%

In this paper Rashba interaction acts within $xy$-plane and current is applied always along $x$.
Instead, our definition of polar angle in spin space 
depends on the wall configuration. 
We define $\theta$ as the angle measured from the easy axis (i.e., $S_{\parallel} = \cos{\theta}$), and  
$\phi$ as the angle in the plane perpendicular to easy axis.
The hard axis is given by $\phi = \frac{\pi}{2}$ 
and so the hard axis component is written as
$S_{\bot} = \sin{\theta} \sin{\phi}$.
Note therefore that ($\theta , \phi$) for N\'eel and Bloch(y) walls below
are different from standard definition measured from $z$-axis.
Lagrangian $L_s$ is thus written in terms of polar angle as 
\eq{
L_s = \inthra \bi{\hbar S \dot{\phi} \lc{\cos{\phi}-1} 
- \frac{J}{2} \lc{(\na \theta)^2 + \sin^2{\theta} (\na \phi)^2} 
- \frac{K S^2}{2} \sin^2{\theta} - \frac{K_{\bot} S^2}{2} \sin^2{\theta} \sin^2{\phi}}, \label{eq:ls1}
} 
for any type of walls considered here.

\subsection{Conduction electron}
The Hamiltonian of the conduction electron is given by the following four terms.
The first term is the free electron part, represented as 
\eq{H_{\rm{e}} =\inthr \cxd \bi{ -\frac{\hbar^2}{2 m} \nabla^2 -\Ef} \cx , \label{freepart}}
where $c$ and $c^\dagger$ are electron annihilation and creation operators,  
$\Ef$ is the Fermi energy and $m$ is the effective mass.
The second is the exchange interaction between electron and local spin, 
\eq{
H_{\rm{ex}} = - \frac{\Delta}{S} \inthr \bm{S}\cdot \lc{\cxd \vet{\sigma} \cx} \label{eq:hex1},
}
where $\Delta$ is the magnitude of the interaction, $\vet{S}$ is the local spin vector and 
$\vet{\sigma}$ are Pauli matrices.
The third one, Rashba spin-orbit interaction,
is given as
\eq{
H_{\rm{so}} = \frac{i \la}{2} 
\inthr \cxd \bi{  (\overleftrightarrow{\nabla}_{x} \pax{y} - \overleftrightarrow{\nabla}_{y} \pax{x} )} \cx ,
\label{rashbapart}}
where $\overleftrightarrow{\nabla}$ acts on both sides and $\la$ is the magnitude of Rashba interaction (the dimensions is [J $\cdot$ m]).
When electric field is applied in the $x$-direction, the momentum, $-i \hbar \ave{\cxd \nax{x} \cx}$, grows,
and Rashba interaction then changes the electron spin direction toward $y$-direction, inducing electron spin precession.

Since we are interested in a response of local spins to applied current,
we introduce the interaction with electric field, given by
(neglecting $O(E^2)$)
\eq{
H_{\rm{EM}} = -e \inthr \bm{A}_{\rm{EM}} \cdot 
\lc{\cxd \vet{v} \cx},
}
where $\bm{A}_{\rm{EM}} = \frac{\bm{E}}{i \Omega} e^{i \Omega t} $ is a $U(1)$ gauge field,
$\bm{E}$ is applied electric field.
Velocity operator is given as
\eq{
v_{\nu} = \frac{- i \hbar \nax{\nu}}{m} +\frac{\la}{\hbar} \lc{\dek{\nu}{y} \pax{x} - \dek{\nu}{x} \pax{y}}.
}
The field $\bm{E}$ is spatially uniform but has finite frequency $\Omega$. 
This frequency is introduced for calculation purpose and is chosen as $\Omega=0$ at the end of calculation, as is usually the case of linear response calculation.

\subsection{Gauge transformation}
The exchange interaction, $H_{\rm{ex}}$, has in general off-diagonal components.
Besides the local ground state of conduction electron varies at each lattice point if magnetization is non-uniform.
In this case, local gauge transformation in spin space~\cite{korenman,gt2} which diagonalizes the exchange interaction is useful.
The spatial change of local spin is then represented by a gauge field,
which is proportional to spatial spin variation, $\partial_{\mu} \vet{S}$.
We consider the case when local spin profile is slowly varying
(called the adiabatic limit), and then gauge field is small.
We thus examine only the first-order contribution of gauge field.
A new electron operator ,$a$, after gauge transformation is defined by use of $2 \times 2$ unitary matrix $U$ 
as
\eq{
\cx = U(x) \ax  \label{eq:sp021},
}
where electron operators here have two spin components like $c\equiv(c_+,c_-)$.
The matrix $U$ is expressed using Pauli matrix as 
\eq{
U(x) = \vet{m}(x) \cdot \vet{\sigma} \label{eq:sp02},
}
where $\vet{m}(x)$ is a vector which characterize the gauge transformation.
We denote the spin easy axis as $\eta$ (e.g., $\eta=x$ for N\'eel wall).
The gauge transformation is defined so that conduction electron spins are polarized along magnetic easy axis $\eta$, 
i.e. to satisfy 
$\frac{\vet{S}(x)}{S} \cdot \lc{\cxd \vet{\sigma} \cx} = \axd \pax{\eta} \ax$.
(The transformation thus differs for different wall configuration.)
The Hamiltonian given by Eqs. (\ref{freepart})(\ref{eq:hex1})(\ref{rashbapart}) 
is written in $a$-electron representation as 
\eq{\notag
H_{\rm e}+H_{\rm ex}+H_{\rm so} 
  = & \int d^3x \Bigg[
 \axdn \bi{- \frac{\hbar^2}{2m} \nabla^2 - \Delta \pax{\eta} -i \la (\pax{y} \nax{x} - \pax{x} \nax{y}) 
 } a 
\ka
& - \frac{i \hbar^2}{m} \sgax{\mu}{\al} \axdn \pax{\al} \nax{\mu} a + \la (\sgax{x}{y} - \sgax{y}{x}) \axdn a \\
& + 2i \la \bi{m^{y} m^{\be} \axdn \pax{\be} \nax{x} a - m^{x} m^{\be} \axdn \pax{\be} \nax{y} a} \Bigg] .}
where gauge field is given as 
\eq{
A_{\mu} \equiv -i U^{\dagger}\nabla_{\mu}U  = \lc{\bm{m} \times \de{\mu} \bm{m}} \cdot \bm{\sigma} \equiv A_{\mu}^{\alpha} {\sigma}^{\alpha}. \label{eq:gaze2}
}
The electromagnetic interaction after gauge transformation is given as
\eq{
H_{EM} = \int d^3x
\frac{- e E_x}{i \Omega} a^\dagger \bi{
\frac{-i\hbar \nabla_{x}}{m} +\frac{\hbar}{m} A_{x} 
-\frac{\la}{\hbar} \overline{\pax{y}}} a,
}
where $\overline{\pax{\mu}} \equiv U^{\dagger} \pax{\mu} U = 2 m^{\mu} \lc{\bm{m} \cdot \vet{\sigma}} - \pax{\mu}$ and
the electric field is applied in $x$-direction.

The total electron Hamiltonian is therefore obtained as 
\eq{H = & \int d^3x 
\Bigg[\notag
\axdn \bi{- \frac{\hbar^2}{2m} \nabla^2 - \Delta \pax{\eta} -i \la (\pax{y} \nax{x} - \pax{x} \nax{y}) 
+ \frac{e E_x}{\Omega} \lc{\frac{\hbar \nax{x}}{m} - \frac{\la}{i \hbar} \pax{y}}} a  \ka
&- \frac{i \hbar^2}{m} \sgax{\mu}{\al} \axdn \pax{\al} \nax{\mu} a + \la (\sgax{x}{y} - \sgax{y}{x}) \axdn a 
+ 2i \la \bi{m^{y} m^{\be} \axdn \pax{\be} \nax{x} a - m^{x} m^{\be} \axdn \pax{\be} \nax{y} a} \\
& - \frac{e E_{x}}{i \Omega} \axdn \bi{\frac{\hbar}{m} \sgax{x}{\al} \pax{\al} - \frac{2 \la}{\hbar} m^{y} \lc{\vet{m} \cdot \vet{\sigma}} } a
\Bigg] \label{eq:ron24}. 
}

\subsection{Effective Hamiltonian}

The current-induced part of the effective Hamiltonian for local spin is directly obtained from Eq. (\ref{eq:ron24}) as
\eq{
H_{\rm{eff}} =& \inthr \Bigg[ \hbar \sgax{\mu}{\al} \js{\mu}{\al} + \la n (\sgax{x}{y} - \sgax{y}{x}) 
 - \frac{2m}{\hbar} \la \bi{m^{y} m^{\be} \js{x}{\be} - m^{x} m^{\be} \js{y}{\be}} 
 - \frac{e E_{x}}{i \Omega} \bi{\frac{\hbar}{m} \sgax{x}{\al} s^{\al} - \frac{2 \la}{\hbar} m^{y} m^{\al} s^{\al} }
 \Bigg].\label{eq:heff35}
}
where electron properties are represented by the following expectation values,
\eq{
n (x) &\equiv \ave{\axd \ax}, \label{eq:nsj01} \\
\spx{\al} (x) &\equiv \ave{\axd \pax{\al} \ax}, \label{eq:nsj02}\\
\js{\mu}{\al} (x) &\equiv - \frac{i \hbar}{2 m} \ave{\axd \overleftrightarrow{\nabla}_{\mu} \pax{\al} \ax \label{eq:nsj03} }. 
}
Here $\left\langle \ \right\rangle$ denotes expectation value evaluated using non-perturbed Hamiltonian $H_a$ defined as
\eq{
H_a \equiv & \int d^3x 
\axd \bi{- \frac{\hbar^2}{2m} \nabla^2 - \Delta \pax{\eta} -i \la (\pax{y} \nax{x} - \pax{x} \nax{y}) 
} \ax , \label{Hadef}
}
and including  to linear order the effect of current,
\eq{
H_{\rm EM}^0 \equiv & \int d^3x 
\axd \bi{\frac{e E_x}{\Omega} \lc{\frac{\hbar \nax{x}}{m} - \frac{\la}{i \hbar} \pax{y}}} \ax .
} 

Current-induced part of the effective Lagrangian is given by
\eq{
L_{{eff}} = \inthr \bi{\hbar \sgax{t}{\al} s^{\al}} - H_{{eff}}. 
}

In calculating the expectation values,
Rashba spin-orbit interaction is treated perturbatively to the  second-order, $\la^2$.
This approximation correspond to assuming $\frac{\la k_f}{\Ef} \ll 1$ with Fermi wave vector $k_F$. 
This expansion with respect to $\la$ is justified by the presence of $\Delta$, in contrast to the non-perturbative nature of unpolarized Rashba system\cite{Inoue}. 
We also assume that the effect of impurities is weak and the electron lifetime is long, i.e. $\frac{1}{\Ef \tau} \ll 1$.
In this case, the term in Eq. (\ref{eq:heff35}) including electric field turns out to be small by a factor of $\frac{1}{\Ef \tau}$ compared with dominant contributions.
We will thus evaluate the effective Hamiltonian given by
\eq{
H_{\rm {eff}} = 
 \inthr \Big[ \hbar \sgax{\mu}{\al} \js{\mu}{\al} + \la n (\sgax{x}{y} - \sgax{y}{x})  
 - \frac{2m}{\hbar} \la \bi{m^{y} m^{\be} \js{x}{\be} - m^{x} m^{\be} \js{y}{\be}} 
\Big].\label{eq:hs41}
}

\section{Bloch(z) case}
In this section, we derive the effective Hamiltonian for the anisotropy configuration of Bloch(z) type, namely,
magnetic easy axis is in the $z$-direction and hard axis is in the $y$-direction. 
\begin{figure}[h]
\begin{minipage}{0.45\textwidth}
\centering
\includegraphics[width=8cm,clip]{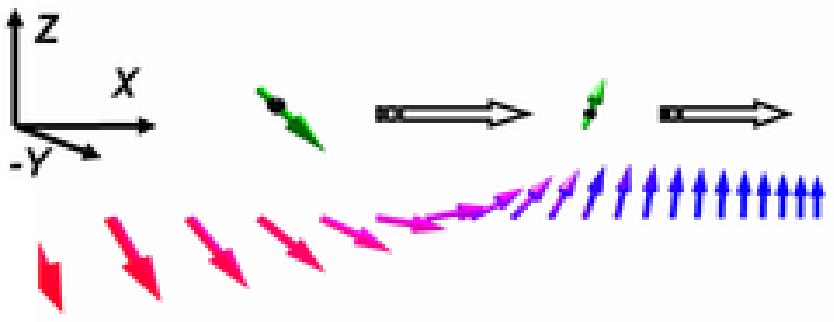} 
\caption{Bloch(z) wall configuration with easy axis along $z$ direction and hard axis along $y$ axis. 
The spin of incoming electron is schematically shown. }
\label{fig:bdwz}
\end{minipage}
\hfill
\begin{minipage}{0.45\textwidth}
\centering
\includegraphics[width=4cm,clip]{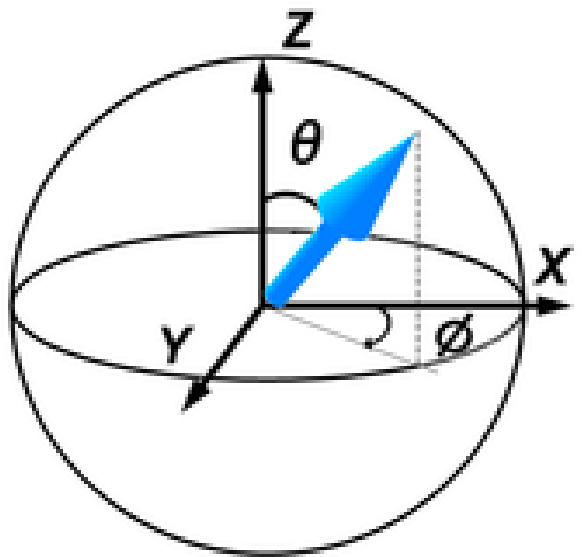}
\caption{Polar coordinates of local spin in the Bloch(z) case.}
\label{fig:spz}
\end{minipage}
\end{figure}%
In this case, domain wall configuration is Bloch(z) wall shown in Fig. \ref{fig:bdwz}.
Polar coordinates are defined by  (as in Fig. \ref{fig:spz})
\begin{align} \vet{S}(x,t)=S(\sin{\theta}\cos{\phi}, \sin{\theta}\sin{\phi}, \cos{\theta}) .\end{align}

In the case of Bloch(z)-type anisotropy, the gauge transformation is defined by 
$\vet{S} (\uxdn \vet{\sigma} U) =S \pax{z}$.
This is satisfied by choosing  
\eq{
\vet{m} = (\sin{\frac{\theta}{2}}\sin{\phi} ,\cos{\frac{\theta}{2}} ,\sin{\frac{\theta}{2}}\cos{\phi} ).
}
Gauge field is then given as
\begin{align}
A_{\mu}=\begin{pmatrix} \sgax{\mu}{x} \\ \sgax{\mu}{y} \\ \sgax{\mu}{z}
\end{pmatrix} =
\frac{1}{2} \begin{pmatrix} -\partial_{\mu}\theta \sin{\phi} - \sin{\theta} \cos{\phi} \partial_{\mu}\phi \\
 \partial_{\mu}\theta\cos{\phi}-\sin{\theta}\sin{\phi}\partial_{\mu}\phi \\
\big(1-\cos{\theta}\big) \partial_{\mu} \phi \end{pmatrix}. 
\end{align}
The unperturbed Hamiltonian, given by choosing $\eta=z$ in Eq. (\ref{Hadef}), reads
\eq{
H_a = & \inthr \axd \bi{-\frac{\hbar^2}{2m} \nabla^2 - \Delta \pax{z} -i \la (\pax{y} \nax{x} - \pax{x} \nax{y})} \ax .
\label{eq:haz1}
} 
This Hamiltonian $H_a$ has the off-diagonal elements in the spin space due to Rashba interaction,
and so we will diagonalize it using a unitary transformation in momentum space, 
$\akd = d_{k}^{\dagger} T_k^{\dagger}$, 
where $d_{k}^{\dagger}$ is a new creation operator and  
$T_k$ is $2 \times 2$ unitary matrix.
The Hamiltonian $H_a$ in momentum space reads 
\begin{align}
H_a = \sum_{k} d_{k}^{\dagger} T_k^{\dagger} \left( \begin{matrix} \Ek -\Delta & -i \la k_{-} \\ i \la k_{+} & \Ek - \Delta \end{matrix} \right) T_k d_k ,\label{eq:haz2}
\end{align}
where $k_{\pm} \equiv k_x \pm i k_y$.
Diagonalization of $H_a$ is done by chosen the unitary matrix $T_k$ as,
\begin{align}
T_k \equiv \frac{1}{\sqrt{A_k^2+\la^2 k^2}} \left( \begin{matrix} A_k & i \la k_{-} \\ -i \la k_{+} & -A_k \end{matrix} \right) ,
\end{align}
where $A_k = \Delta + Z_k, Z_k = \sqrt{\Delta^2+\la^2 k^2}$.
The result is 
\begin{align}
H_a = \sum_{k} d_k^{\dagger} \left( \begin{matrix} \Ek - Z_k & 0 \\ 0 & \Ek + Z_k \end{matrix} \right) d_k \label{eq:mha65}.
\end{align}
After diagonalization, the energy of conduction electrons is spin split as
\eq{\E_{k,\sigma} = \frac{\hbar k^2}{2 m} - \Ef - \sigma Z(k), \label{eq:ekz}}
where $\sigma = (+,-)$ is spin polarization.

\begin{figure}[htb]
\centering
\includegraphics[width=0.7\linewidth,clip]{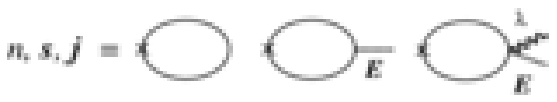} 
\caption{Diagramatic representation of electron density ($n$), spin density ($s^\alpha$) and spin current ($j_{{\rm s},i}^\alpha$) at the linear order in the applied electric field, $\bm{E}$. 
Solid line represents electron Green function with Rashba interaction included. 
Vertex denoted by $\times$ is $1$, $\sigma^\alpha$ and $k_i\sigma^\alpha$, for $n$, $s^\alpha$ and $j_{{\rm s},i}^\alpha$, respectively.
}
\label{figdiag}
\end{figure}

Now, let us estimate expectation values.
The electron density $n$ (Fig. \ref{figdiag}) is given as 
\eq{\notag
n =& - \sum_{k,\ome} i \tr{\ti{G}_{k,\w}}^{<} \ka 
& + \lio \sum_{k,\ome} \frac{e E_x}{ \Omega} \frac{\hbar^2 k_x}{m} \trles{\ti{G}_{k,\w} \ti{G}_{k,\Ome}} \\ 
& + \lio \sum_{k,\ome} \frac{e E_x}{ \Omega} \la \trles{\ti{G}_{k,\w} \tpax{y} \ti{G}_{k,\Ome}}, \label{eq:nd3}
}
where tr is a trace in spin space and
$\tpax{\al}\equiv T_k^{\dagger} \pax{\al} T_k$.
Keldysh Green function~\cite{keld,smith} with Rashba interaction included, $\ti{G}_{k,\w}^{<}$, is
defined as
\begin{align}
\notag
\ti{G}_{k,\w}^{<} \equiv & i \ave{d_{k,\ome}^{\dagger} d_{k,\ome}}   \\
= & \begin{pmatrix} g_{k,\w,\up}^{<} & 0 \\
0 & \ti{g}_{k,\w,\dw}^{<} 
\end{pmatrix},
\end{align}
where component $\tgwl{k,\ome}{\sigma}$ is given as
\eq{
\tgwl{\ome}{\sigma} = 2 \pi i f( \ome ) \dex{\hbar \ome}{\E_{k,\sigma}},
}
 $f( \ome )\equiv \frac{1}{e^{\beta\omega}+1}$  being Fermi distribution function ($\beta$ is inverse temperature).
Summation over $k$ is carried out in two-dimensions by replacing by energy integration, 
$\sum_{k} = \frac{V m}{2 \pi \hbar^2} \int d\E$.
Products of Keldysh Green functions is calculated using  relations
\eq{
\ti{G}_{k,\w}^{<} \equiv f( \w ) \bi{\ti{G}_{k,\w}^{a} - \ti{G}_{k,\w}^{r}} ,
}
where $\ti{G}_{k,\w}^{r}$ ($\ti{G}_{k,\w}^{a}$)
is retarded (advanced) Green function, 
and
\eq{\notag
\lesbi{G_{k,\w} G_{k',\Ome}} =& G_{k,\ome}^{r} {G}_{k',\Ome}^{<} + G_{k,\ome}^{<} {G}_{k',\Ome}^{a} \\
=& G_{k,\ome}^{r} {G}_{k',\Ome}^{a} \lc{f( \Ome ) - f( \w )} - G_{k,\ome}^{r} {G}_{k',\Ome}^{r} + G_{k,\ome}^{a} {G}_{k',\Ome}^{a}
.}
Expanding with respect to $\la$ to the second order, we obtain the density as
\eq{
n = \frac{m}{\pi \hbar^2} \Ef + \frac{m^2 \la^2}{\pi \hbar^4}.
}
We see that there is no effect from the applied electric field here.

The electron spin density $s^{\g}$ is similarly calculated as
\eq{\notag
s^{\g} =& - \sum_{k,\ome} i \trles{\tpax{\g} \ti{G}_{k,\w}} \ka 
& + \lio \sum_{k,\ome} \frac{e E_x \hbar^2 k_x}{ \Omega m} \trles{\tpax{\g} \ti{G}_{k,\w} \ti{G}_{k,\Ome}} \\ 
& + \lio \sum_{k,\ome} \frac{e E_x \la}{\Omega} \trles{\tpax{\g} \ti{G}_{k,\w} \tpax{y} \ti{G}_{k,\Ome}}. \label{eq:spin3} 
}
The result is 
\eq{
s^x =& - \frac{m e E_x \la}{2 \pi \hbar^2 \Delta} \nonumber\\
s^y =&  \frac{m e E_x \tau \la}{\pi \hbar^2} \nonumber\\
s^z =& \frac{m \Delta}{\pi \hbar^2} \label{eq:spz3}.
}
We see that the electric field induces perpendicular components $s^x$ and $s^y$, but $\abs{s^y} \gg \abs{s^x}$ since $\frac{1}{\Delta \tau} \ll 1$. 
We will thus approximate $\abs{s^x}\simeq0$.
The $z$-component of spin in Eq. (\ref{eq:spz3}) is the adiabatic contribution, which is not affected by applied field. 

Spin current, $\js{\mu}{\al}$, is estimated using
\eq{\notag
i \ave{\akd k_{\mu} \pax{\al} \ak} =& \sum_{k,\ome} k_{\mu} \trles{\tpax{\al} \ti{G}_{k,\w}}  \ka
& - \lio \sum_{k,\ome} k_\mu \frac{e E_x \hbar^2 k_x}{i \Omega m} \trles{\tpax{\al} \ti{G}_{k,\w} \ti{G}_{k,\Ome}}  \\
& - \lio \sum_{k,\ome} k_\mu \frac{e E_x \la}{i \Omega} \trles{\tpax{\al} \ti{G}_{k,\w} \tpax{y} \ti{G}_{k,\Ome}}, \label{eq:spcr3}  
}
as
\eq{
\js{y}{x} =&  \frac{m}{\pi \hbar^3} \la \Ef \nonumber \\
\js{x}{y} =& -\frac{m}{\pi \hbar^3} \la \Ef \nonumber \\
\js{x}{z} =& - \frac{e E_x \tau}{\pi \hbar} \bi{\Delta - \frac{m \la^2}{2 \Delta \hbar^2} \Ef}.
}
We see that spin current are generated in $x$ and $y$ direction by the Rashba interaction without electric field and that
$\js{y}{x}$ is equal to $- \js{x}{y}$. 
This is due to the symmetry of Rashba spin orbit interaction.
In contrast, $\js{x}{z}$ is induced by applied electric field.
We define the current density (divided by $e$) and electron density without spin-orbit interaction as 
($\sigma$ is Boltzmann conductivity) 
\eq{
J \equiv - \frac{e E_x \tau \Ef}{\pi \hbar} = \frac{\sigma}{e} E_x\label{eq:cr0},
}
and 
\eq{
n_0 \equiv \frac{\Ef m}{\pi \hbar^2} \label{eq:nd0} .
}
In term of these parameters,
the above result reads 
\begin{align}
n =& n_0 \bi{1 + \frac{m \la^2}{2 \hbar^2 \Ef}} \nonumber \\
\lc{s^x, s^y, s^z} =& \lc{ 0, - \frac{m \la}{\hbar \Ef} J, n_0 \frac{\Delta}{\Ef}}\nonumber  \\
\lc{\js{x}{y}, \js{x}{z}, \js{y}{x}} =& \lc{- \frac{\la}{\hbar} n_0 , \bi{\frac{\Delta}{\Ef} - \frac{m \la^2}{2 \Delta \hbar^2}} J, \frac{\la}{\hbar} n_0}. \label{eq:js44}
\end{align}
Other components of spin current vanish.

The effective Hamiltonian for Bloch(z) case is therefore obtained from Eq. (\ref{eq:hs41}) as 
\eq{\notag
H_{\rm {eff}} =& \inthr \Bigg[ \frac{\hbar}{2} \lc{1- \cos{\theta}} \lc{\de{x}\phi} \js{x}{z}  \\
&-\frac{\la m}{\hbar} \bi{\js{x}{z} \sin{\theta} \sin{\phi} + \lc{1- \cos{\theta}} \lc{\js{x}{y} - \sin{\phi} \js{y}{x} } } \Bigg] \label{eq:ey128}.
 }
We see that applied current induces 
$\js{x}{z}$ (Eq. (\ref{eq:js44})) and this induces when coupled with Rashba interaction an effective magnetic field in $y$-direction as indicated by 
the second term. 
The first term of Eq. (\ref{eq:ey128}) represents standard spin transfer torque (with current modified by Rashba interaction).
The third term is independent of applied current and is a modification of magnetic anisotropy by Rashba interaction.

\section{N\'eel(x) case}
\begin{figure}[h]
\begin{minipage}{0.45\textwidth}
\centering
\includegraphics[width=7cm,clip]{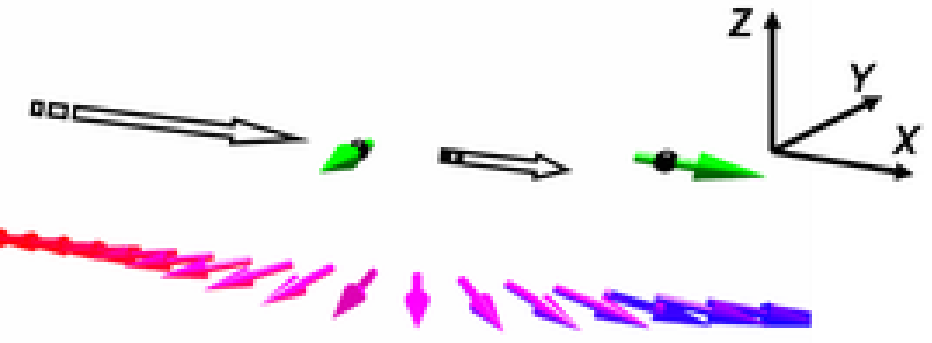} 
\caption{N\'eel(x) wall configuration.}
\label{fig:dwx1}
\end{minipage}
\hfill
\begin{minipage}{0.45\textwidth}
\centering
\includegraphics[width=4cm,clip]{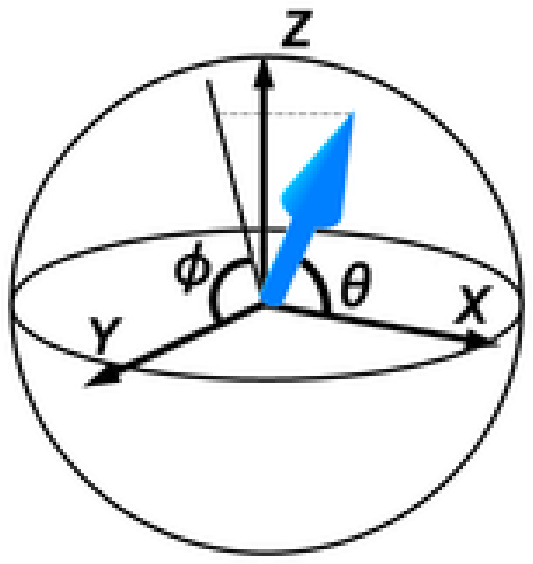}
\caption{Definition of polar coordinates for N\'eel(x) case.}
\label{fig:spx}
\end{minipage}
\end{figure}%
In this section, we consider a case of N\'eel wall 
realized when magnetic easy axis and hard axis are in the $x$ and $z$ direction, respectively (Fig. \ref{fig:dwx1}).
Polar coordinates are defined differently from standard definition as (Fig. \ref{fig:spx}) 
\eq{
\vet{S} = S (\cos{\theta} ,\sin{\theta}\cos{\phi} ,\sin{\theta}\sin{\phi}). \label{eq:s62}
}

Derivation of effective Hamiltonian is done similarly to Bloch(z) case.
Difference is in definition of gauge transformation, 
$\vet{S} (U^{\dagger} \vet{\sigma} U) =S \pax{x}$.
Vector $\vet{m}$ is accordingly chosen as
\eq{
\vet{m} = (\cos{\frac{\theta}{2}} ,\sin{\frac{\theta}{2}}\cos{\phi} ,\sin{\frac{\theta}{2}}\sin{\phi}),
}
and $\sgax{\mu}{\al}$ is given as
\begin{align}
A_{\mu}=
\frac{1}{2}\left( \begin{matrix}
(1-\cos{\theta}) \de{\mu}\phi \\
-\sin{\phi} \de{\mu}\theta - \sin{\theta} \cos{\phi} \de{\mu} \phi \\
 \de{\mu}\theta \cos{\phi} - \sin{\theta} \sin{\phi} \de{\mu}\phi
\end{matrix} \right) .
\end{align}
Hamiltonian $H_a$ is also different from Bloch(z) case since uniform spin polarization is now along $x$-direction.
It is given as
\begin{align}
H_a = \sum_k 
\akd \left( \begin{matrix} \Ek & -b_- \\ -b_+ & \Ek \end{matrix}  \right) \ak ,
\label{eq:hax132}
\end{align}
where $b_{\pm} \equiv (\Delta \mp i \la k_{\pm})$.
The diagonalization of $H_a$ is carried out as $\akd = \dkd T_k^{\dagger}$ where  
\begin{align}
T \equiv \frac{1}{\sqrt{2}Z_k} \left(\begin{matrix} Z_k & b_- \\ b_+ & -Z_k \end{matrix}\right) ,
\end{align}
where 
\eq{
Z_k = \sqrt{\Delta^2 + 2 \la \Delta k_y + \la^2 k^2} .
}
The Hamiltonian after diagonalization reads
\begin{align}
H_a = \sum_{k} \dkd \left( \begin{matrix} \Ek - Z_k & 0 \\ 0 & \Ek + Z_k \end{matrix} \right) \dk \label{eq:hax1}.
\end{align}
The expectation values are calculated similarly to Bloch(z) case and we obtain 
\eq{
n =& n_0\\
\lc{s^x, s^y} =& \lc{n_0 \bi{\frac{\Delta}{\Ef} - \frac{\la^2}{2 \Delta}},
- \frac{m \la}{\Ef \hbar} J} \\
\lc{\js{x}{x} ,\js{x}{y} ,\js{y}{x} } =& \lc{\frac{\Delta}{\Ef} J,
- \frac{\la}{\hbar} n_0,
\frac{\la}{\hbar} n_0 },
}
where other components of spin and spin current vanish.

The current-induced effective Hamiltonian is then obtained as 
\eq{\notag
H_{\rm eff} =& \inthr \frac{\hbar}{2} \lc{1-\cos{\theta}} \lc{\de{x} \phi} \js{x}{x} \\ 
& -\frac{m}{\hbar} \la \bi{\js{x}{x} \sin{\theta} \cos{\phi} +\lc{1-\cos{\theta}} \lc{\js{x}{y} \cos^2{\phi} - \js{y}{x}} \label{eq:he195}} .
 }
The applied current coupled with spin-orbit interaction induces an effective magnetic field
along hard, i.e., $y$ axis 
(second term of Eq. (\ref{eq:he195})), as in  the Bloch(z) case.

\section{Bloch(y) case}
\begin{figure}[h]
\begin{minipage}{0.45\textwidth}
\centering
\includegraphics[width=7cm,clip]{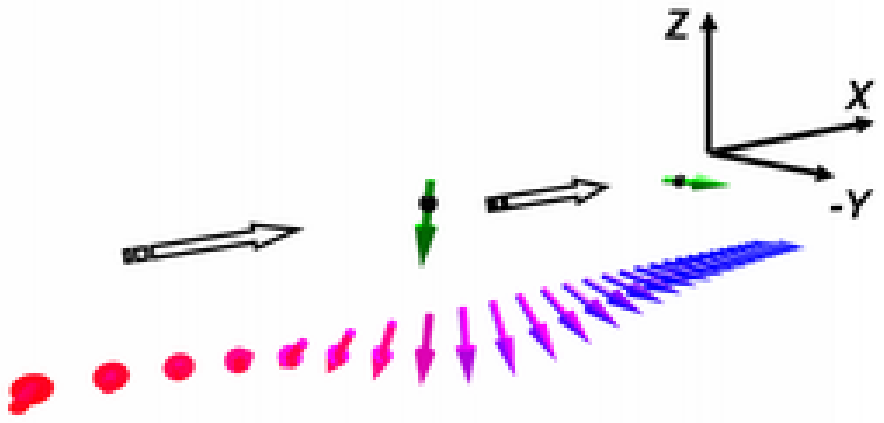} 
\caption{Bloch(y) wall configuration.}
\label{fig:dwy1}
\end{minipage}
\hfill
\begin{minipage}{0.45\textwidth}
\centering
\includegraphics[width=4cm,clip]{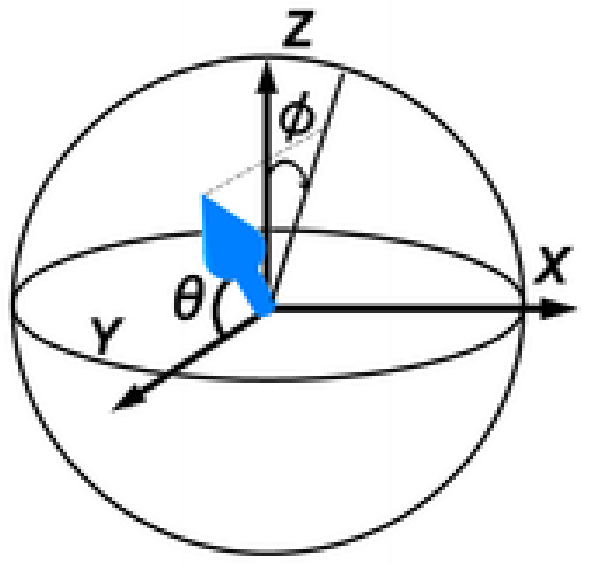}\\
\caption{Polar coordinates for Bloch(y) configuration. }
\label{fig:spy}
\end{minipage}
\end{figure}%
Finally, we consider a case of 
magnetic easy axis and hard axis along $y$ and $x$ direction, respectively.
Domain wall in this case is Bloch(y) wall (Fig. \ref{fig:dwy1}).
Polar coordinates are define as (Fig. \ref{fig:spy}) 
\eq{
\vet{S} = S (\sin{\theta}\sin{\phi} , \cos{\theta} , \sin{\theta} \cos{\phi}).
}
Gauge transformation is given by
 $\vet{S} (U^\dagger \vet{\sigma} U) =S \pax{y}$,
and vector $\vet{m}$ consequently becomes 
\eq{
\vet{m} = (\sin{\frac{\theta}{2}}\sin{\phi} ,\cos{\frac{\theta}{2}} ,\sin{\frac{\theta}{2}}\cos{\phi} ),
}
and gauge field is 
\begin{align}
A_{\mu}=
\frac{1}{2}\left( \begin{matrix}
 \de{\mu}\theta \cos{\phi} - \sin{\theta} \sin{\phi} \de{\mu}\phi \\
(1-\cos{\theta}) \de{\mu}\phi \\
-\sin{\phi} \de{\mu}\theta - \sin{\theta} \cos{\phi} \de{\mu} \phi \\
\end{matrix} \right) .
\end{align}
Hamiltonian $H_a$ is given as
\begin{align}
H_a = \sum_{k} \akd 
\left(\begin{matrix} \Ek & -i b_- \\ -i b_+ & \Ek \end{matrix} \right) \ak ,
\end{align}
where $b_{\pm} = (\Delta \mp \la k_{\pm})$ (is different from the N\'eel case).
Hamiltonian $H_a$ is diagonalized as
\begin{align}
H_a = \sum_{k} \dkd \left( \begin{matrix} \Ek - Z_k & 0 \\ 0 & \Ek + Z_k \end{matrix} \right) \dk \label{eq:hay01},
\end{align}
by defining 
$\akd = d^{\dagger} T_k^{\dagger}$ 
with
\begin{align}
T \equiv \frac{1}{\sqrt{2}Z_k} \left( \begin{matrix} -Z_k & i b_{-} \\ -i b_{+} & Z_k \end{matrix} \right) ,
\end{align}
where 
\eq{
Z_k \equiv \sqrt{\Delta^2 - 2 \la \Delta k_x + \la^2 k^2} .
}
Expectation values which appears in the effective Hamiltonian are estimated as 
\eq{
n =& n_0 \bi{1 - \frac{3 m}{\Ef \hbar^2} \la^2} \nonumber \\
s^y =& n_0 \bi{\frac{\Delta}{\Ef} + \frac{m \la^2}{\Delta \hbar^2}} \la J \nonumber \\
\lc{\js{x}{y}, \js{y}{x}} =& \lc{- n_0 \frac{\la}{\hbar} + J \bi{\frac{\Delta}{\Ef} + \frac{m \la^2}{\Ef \hbar^2} +\frac{2 m \la^2}{\Delta \hbar^2}},
n_0 \frac{\la}{\hbar} - \frac{m \la^2}{\Delta \hbar^2} J }.
}
The effective Hamiltonian is therefore obtained as
\eq{
H_{\rm{eff}} = & \inthr \left[
\frac{1}{2} (1-\cos{\theta}) \lc{\de{x} \phi} \bi{\hbar \js{x}{y} +\la n} \right. \nonumber\\
&\left.
-\frac{m \la}{\hbar} \bi{\lc{1 + \cos{\theta}} \js{x}{y} - \lc{1 - \cos{\theta}} \js{y}{x} \sin^2{\phi}  } \right] \label{hey2}.
}

\section{Analysis of effective Hamiltonian}
The effective Hamiltonian representing the effect of current  
obtained above is summarized as follows.\\
Bloch(z) case:
\eq{\notag
H_{\rm{eff}} =& \int d^3x \left[ \frac{\hbar}{2} \lc{1- \cos{\theta}} \lc{\de{x}\phi} \bi{\frac{\Delta}{\Ef} - \frac{m}{2 \Delta \hbar^2} \la^2} J 
-\frac{\la m \Delta}{\Ef \hbar} \es{y} J \right. \\
& \left. + \frac{\la^2 m}{\hbar^2} n_0 \lc{1 - \es{z} } \lc{1 + \sin{\phi}} \right].
}
Neel(x) case:
\eq{\notag
H_{\rm{eff}} =& \int d^3x \left[ \frac{\hbar}{2} \lc{1- \cos{\theta}} \lc{\de{x}\phi} \frac{\Delta}{\Ef} J 
-\frac{\la m \Delta}{\Ef \hbar} \es{y} J \right. \\
& \left.+ \frac{m}{\hbar} \la^2 \lc{1 - \es{x}} n_0 \lc{1+ \cos^2{\phi}} \right].
}
Bloch(y) case:
\eq{\notag
H_{\rm{eff}} =& \int d^3x \left[ \frac{\hbar}{2} \lc{1- \cos{\theta}} \lc{\de{x}\phi} \bi{\frac{\Delta}{\Ef} + \frac{m}{\Ef \hbar^2} \la^2 + \frac{2m}{\Delta \hbar^2} \la^2} J 
-\frac{\la m \Delta}{\Ef \hbar} \es{y} J \right. \\
& \left. + \frac{m}{\hbar^2} \la^2 n_0 \bi{\es{y} +\sin^2{\phi} \lc{1-\es{y}}} \right]. \label{heffbly}
}
Here $\es{\al}$ is local spin component in $\al$-direction.
The first term of each effective Hamiltonian represents spin transfer torque, which is enhanced by Rashba interaction in Bloch(z) and Bloch(y) cases at the second order in spin-orbit interaction. The enhancement is thus small but  independent of sign of $\la$.
The second term of the effective Hamiltonian, $\es{y} \la \frac{\Delta}{\Ef} J$, indicates that the effective magnetic field arises in the $y$-direction in all three cases.
This is a result of spin current induced by Rashba interaction and applied current.
The last term of each Hamiltonian exists without electric field and thus represents magnetic anisotropy modified by Rashba interaction.
This change of magnetic anisotropy is  at the  second-order in  $\la$ and is small.
Furthermore, in reality, this static contribution should be contained already in the anisotropy parameters $K , K_{\bot}$. We therefore do not take it into account in the following analysis.

The effects of Rashba interaction are summarized in Table. \ref{dwsoi1}.
\begin{table}[h]
\centering
\begin{tabular}{lccc} \hline \hline
 & \hspace{0.04 \linewidth} Bloch(z) \hspace{0.04 \linewidth}&N\'eel(x)\hspace{0.04 \linewidth}&Bloch(y) \hspace{0.04 \linewidth}\\ \hline
Enhancement of spin transfer torque & $O(\la^2$) & $\times$& $O(\la^2$) \\ \hline  
Effective magnetic field & $O(\la$)  & $O(\la$)& $O(\la$) \\ \hline
Change of magnetic anisotropy & $O(\la^2$)& $O(\la^2$)& $O(\la^2$)\\ \hline \hline
\end{tabular}
\caption{
Summary of effects of Rashba interaction for three configurations represented by the order of the effects in $\la$. 
Symbol $\times$ denotes the absence of the effect.}
\label{dwsoi1}
\end{table}%

\section{Equation of motion of Domain Wall}
Let us discuss how Rashba spin-orbit interaction affects current-induced domain wall motion based on the effective Hamiltonian we derived.
For this Lagrangian formalization is convenient.
We consider rigid planar domain wall.
The Lagrangian for domain wall is then obtained as $L = \dot{X} \phi - (H_S+H_{\rm{eff}})|_{X,\phi}$ where $X$ is domain wall position, $\phi$ is local spin angle in magnetic easy plane and 
$(H_S+H_{\rm{eff}})|_{X,\phi}$ is effective spin Hamiltonian evaluated for domain wall configuration\cite{begt2}.
Explicitly, domain wall Lagrangian is given  as follows.\\
Bloch(z) wall:
\eq{\notag
\ti{L} =& \bi{\phi \ti{\dot{X}} - \sin^2{\phi}} + \cos{\phi} \ti{\dot{\phi}} 
- \frac{\pi s^y}{2} \bi{\ti{\dot{X}} \cos{\phi} - \sin{\phi} \ti{\dot{\phi}}}  \\ 
&- s^z \ti{X}  \ti{\dot{\phi}} - \phi \bi{\frac{\Delta}{\Ef} - \frac{\hbar^2}{2 \Delta m \ell^2} \tla^2} \ti{J} 
+ \pi \tla \sin{\phi} \frac{\Delta}{\Ef} \ti{J}.
\label{eq:lez274}}
Neel(x) wall:
\eq{
\ti{L} = \bi{\phi \ti{\dot{X}} -\sin^2{\phi}} - s^x \ti{X} \ti{\dot{\phi}} 
 + \frac{\pi s^y}{2} \bi{\ti{\dot{X}} \sin{\phi} + \cos{\phi} \ti{\dot{\phi}}} 
 - \phi \frac{\Delta}{\Ef} \ti{J} + \pi \tla \cos{\phi} \frac{\Delta}{\Ef} \ti{J}. \label{eq:lex275} 
}
Bloch(y) wall:
\eq{\notag
\ti{L} =& \bi{\phi \ti{\dot{X}} -\sin^2{\phi}} 
- s^y \frac{X}{\ell} \ti{\dot{\phi}} \ka
& - \phi \bi{\frac{\Delta}{\Ef} + \frac{\hbar^2 \tla^2}{m \ell^2 \Ef} + \frac{2 \hbar^2 \tla^2}{m \ell^2 \Delta}} \ti{J} 
- 2 \tla \frac{X(t)}{\ell} \frac{\Delta}{\Ef} \ti{J}.
}
Here we introduced following dimensionless parameters ,
\eq{
&\tilde{L} \equiv \frac{\ell}{\hbar N S v_c} L, \hspace{1truecm}
\ti{X} \equiv \frac{X}{\ell}, \hspace{1truecm} 
 \tilde{\dot{X}} \equiv \frac{\dot{X}}{v_c}, \hspace{1truecm} 
\tilde{\dot{\phi}} \equiv \frac{\ell}{v_c}\dot{\phi}, 
\notag \\
&
 \ti{J} \equiv  \frac{J}{v_c}, \hspace{1truecm}
\hspace{1truecm} 
\tilde{\la} \equiv \frac{m \ell}{\hbar^2} \la.
\notag
}
where $N$ is a number of local spin in the domain wall and $\ell$ is domain wall thickness given as $\ell \equiv \sqrt{\frac{J}{K}}$. 
(Dimensionless time is $\ti{t} \equiv \frac{v_c}{\ell}t$).
Here a velocity $v_c \equiv \frac{K_{\bot} S \ell}{2 \hbar}$ correspond to drift velocity of electron at intrinsic threshold current without Rashba interaction\cite{begt2}.

Equations of motion for domain wall is obtained taking account of dissipation as,($Q = \ti{X}, \phi $)
\eq{
\devx{\tilde{t}} \dev{\tilde{L}}{\dot{Q}} - \dev{\tilde{L}}{Q} = - \dev{\tilde{W_s}}{\dot{Q}}  \label{eq:eqm},
}
where $\tilde{W_s}$ is dimensionless dissipation function written as 
$\tilde{W_s} = \frac{\al}{2} \bi{\tilde{\dot{X}}^2 + \tilde{\dot{\phi}}^2 }$\cite{TKS08}.

In deriving the equation of motion, we neglect contribution  of conduction electron density ($s^{\g} \ll 1$) since they turns out to be small in actual situations.
The equation of motion is obtained as follows.\\
Bloch(z) wall:
\eq{
\tdt{X}- \al \ti{\dot{\phi}} =& \sin{2 \phi} + \bi{\frac{\Delta}{\Ef} - \frac{\hbar^2}{2 \Delta m \ell^2} \tla^2} 
\ti{J} - \pi \tla \cos{\phi} \frac{\Delta}{\Ef} \ti{J} \label{eq:eqm291},
}
\eq{
\tdt{\phi} + \al \tdt{X} = 0 \label{eq:eqm292},
}
where time evolution of $\phi$ reduces to a single equation of 
\eq{
\tdt{\phi} = -\frac{\al}{1 + \al^2} \bi{\sin{2 \phi} + \bi{\frac{\Delta}{\Ef} 
- \frac{\hbar^2}{2 \Delta m \ell^2} \tla^2} \ti{J} - \pi \tla \cos{\phi} \frac{\Delta}{\Ef} \ti{J} } \label{eq:phyz}.
}
Neel(x) wall: 
\eq{
\tdt{X} - \al \tdt{\phi} =& \sin{2 \phi} + \frac{\Delta}{\Ef} \ti{J} + \pi \tla \sin{\phi} \frac{\Delta}{\Ef} \ti{J},
\label{eq:eqmxx} }
\eq{
\tdt{\phi}+ \al \tdt{X} =& 0 \label{eq:eqx104},
}
which result in
\eq{
\tdt{\phi} =& - \frac{\al}{1 + \al^2} \bi{\sin{2 \phi} + \frac{\Delta}{\Ef} \ti{J} + \pi \tla \sin{\phi} \frac{\Delta}{\Ef} \ti{J} } \label{eq:phyx}.
}
Bloch(y) wall: 
\eq{
\tdt{X} - \al \tdt{\phi} =& \sin{2 \phi} + \bi{\frac{\Delta}{\Ef} + \frac{\hbar^2 \tla^2}{m \ell^2 \Ef} + \frac{2 \hbar^2 \tla^2}{m \ell^2 \Delta}} \ti{J} \label{eq:eyy106},
}
\eq{
\tdt{\phi}+ \al \tdt{X} =& - \frac{2 \tla \Delta}{\Ef} \ti{J} \label{eq:eyq302},
}
and the time evolution of $\phi$ is obtained as
\eq{
\tdt{\phi} =& -\frac{\al}{1 + \al^2} \bi{
\sin{2 \phi} 
+ \bi{\frac{\Delta}{\Ef} + \frac{\hbar^2 \tla^2}{m \ell^2 \Ef} + \frac{2 \hbar^2 \tla^2}{m \ell^2 \Delta}} \ti{J} 
+ \frac{2 \hbar \tla^2}{m \ell v_c} n_0 \sin{2 \phi}  \frac{X}{\ell} 
+ \frac{2}{\al} \frac{\tla \Delta}{\Ef} \ti{J} \label{eq:phyy}
}.}

We here see a large difference between Bloch(y) and other cases.
In fact, the angle $\phi$ in Bloch(y) wall is directly driven by current and Rashba interaction 
as indicated by the right-side of Eq. (\ref{eq:eyq302}).
Such effect of current has been known as $\beta$ terms in the case of electron spin relaxation due to random spin~\cite{Li04,Zhang04}.
In the present Rashba case, the parameter $\be$ is then given by 
\eq{
\be = - \frac{2 \tla \Delta}{\Ef}, \label{eq:beta67}
}
for Bloch(y) 
($\be = 0$ for Bloch(z) and Neel(x) walls).
This result can be explained by noting that $\be$ terms is effectively equivalent to 
an external magnetic magnetic field and that Rashba interaction induces an effective magnetic field, which coinsides with the easy axis for Bloch(y) wall. (Eq. (\ref{heffbly}).
We will see below that this coefficient $\be$ is quite large even assuming standard semiconducting systems and reduces much the threshold current and enhances the wall velocity.
The effect would be even stronger if the systems has giant Rashba effect as realized in metallic surfaces\cite{Ast07}. 

The effect of effective magnetic field due to Rashba interaction in Bloch(z) and Neel(x) cases
is to induce anisotropy within $\phi$-plane as seen as $\cos\phi$ and $\sin \phi$ terms in Eqs. (\ref{eq:eqm291}) and (\ref{eq:eqmxx}), respectively.
This anisotropy energy turns out to drive stepwise wall motion at low current.

We also see from Eqs. (\ref{eq:eqm291}) and (\ref{eq:eyy106}) that spin transfer torque effect 
(represented by the second term of right-hand side)
is enhanced by Rashba interaction,
at the second oder of $\la$
in Bloch(z) and Bloch(y) cases, but not in Neel(x) case.
Numerically, these second order effects are negligibly small as we will show below.

\section{Domain wall dynamics}
We first note that domain wall velocity $\tdt{X}$ is closely related to $\tdt{\phi}$~\cite{begt2}.
In fact, from Eqs. (\ref{eq:eqm292}),(\ref{eq:eqx104}),(\ref{eq:eyq302}), we see that 
\eq{
\tdt{X} = - \frac{1}{\al} \tdt{\phi}
 \;\;\; \mbox{\rm  (Bloch(z) and Neel(x) wall)},\label{eq:beta35}
}
and 
\eq{
\tdt{X} = - \frac{1}{\al} \tdt{\phi} - \frac{2 \tla \Delta}{\al \Ef} \ti{J}  \;\;\; \mbox{\rm (Bloch(y) wall)}. \label{eq:beta22}
}

Here, we show numerical results based on the equation of motion.
Parameter are chosen as 
\eq{\notag
a = 5.65 \times 10^{-10} {\rm m}, \hspace{0.5truecm}
m = 0.05 {m_e} , \hspace{0.5truecm}
m_e = 9.11 \times 10^{-31} {\rm kg},\\ \notag
v_c = 163 {\rm m/s}, \hspace{0.5truecm}
\Ef = 80 {\rm m eV}, \hspace{0.5truecm}
\Delta = 40 {\rm m eV}, \hspace{0.5truecm}
\ell = 50 {\rm nm}, \hspace{0.5truecm}
\al = 0.01.
}
to simulate actual semiconductor systems~\cite{yamano}.
(We thus have $\frac{\hbar^2}{m\ell^2 \Ef}\simeq 0.007$, and so the second order contribution from Rashba interaction is very small like $\tilde\lambda^2 \frac{\hbar^2}{m\ell^2 \Ef}\simeq 0.7\times 10^{-4}$ for $\tilde\lambda=0.1$.)
We calculated the domain wall position $\ti{X}(t)$ after current is applied at $t = 0$.
\begin{figure}[h]
	\centering
\includegraphics[width=16cm,clip] {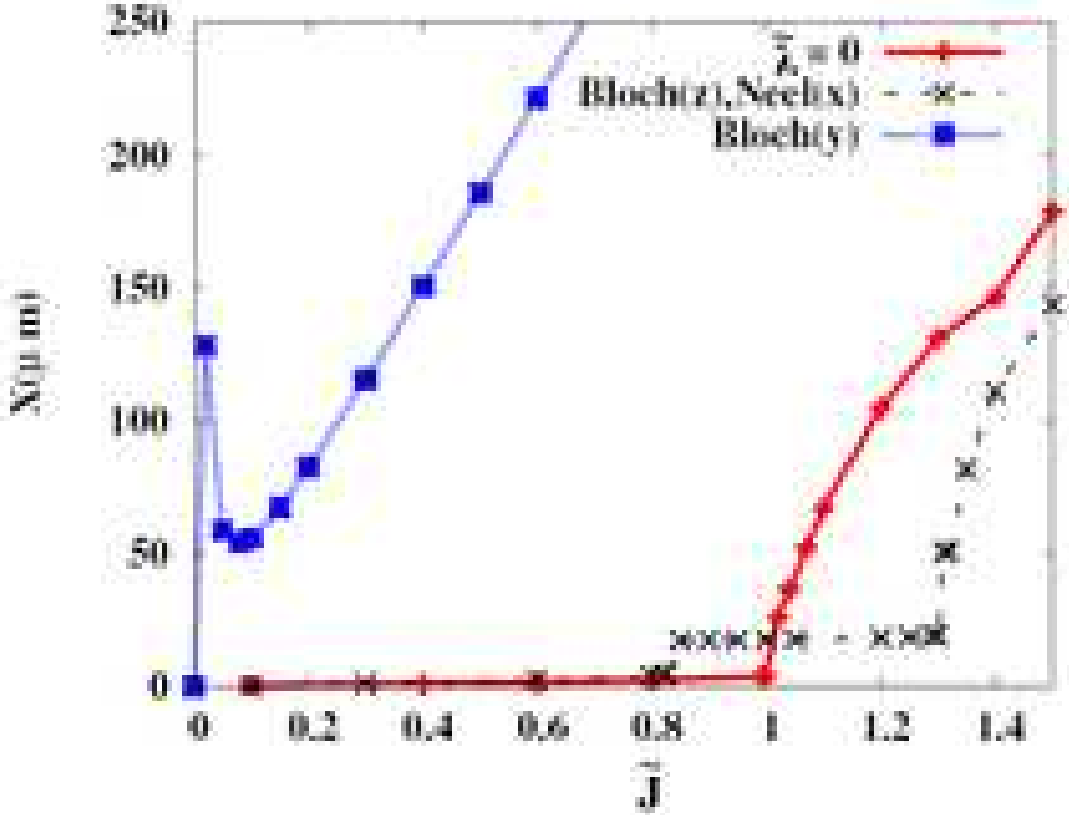} \\ 
	\caption{Wall position at $t=1\mu$s as function of dimensionless current, $\tilde{J}=J/v_c$. 
Solid line represents the case $\tilde\lambda=0$, and 
dashed and dotted lines represent the case  $\tilde\lambda=0.1$ for Bloch(z), N\'eel(x) and Bloch(y) walls, respectively.
}
\label{fig:grawall01}
\end{figure}%
Fig. \ref{fig:grawall01} shows the wall position at $t = 1 \mu m$ after current is applied.
(Current is normalized by $v_c$, the electron drift velocity at intrinsic threshold.)
The case without Rashba interaction is shown as solid line.
We see here the intrinsic pinning due to hard axis anisotropy\cite{begt2}, since we do not consider $\beta$ term of non-Rashba origin.
Wall motion in the presence of Rashba interaction with $\tilde\lambda=0.1$ is plotted by lines
 marked by $\times$ for Bloch(z) and Neel(x) and $\ast$ 
for Bloch(y) walls.
Bloch(z) and Neel(x) walls behaves essentially the same.

\subsection{Bloch(y) wall}

\begin{figure}[h]
  \centering
 \includegraphics[width=14cm,clip]{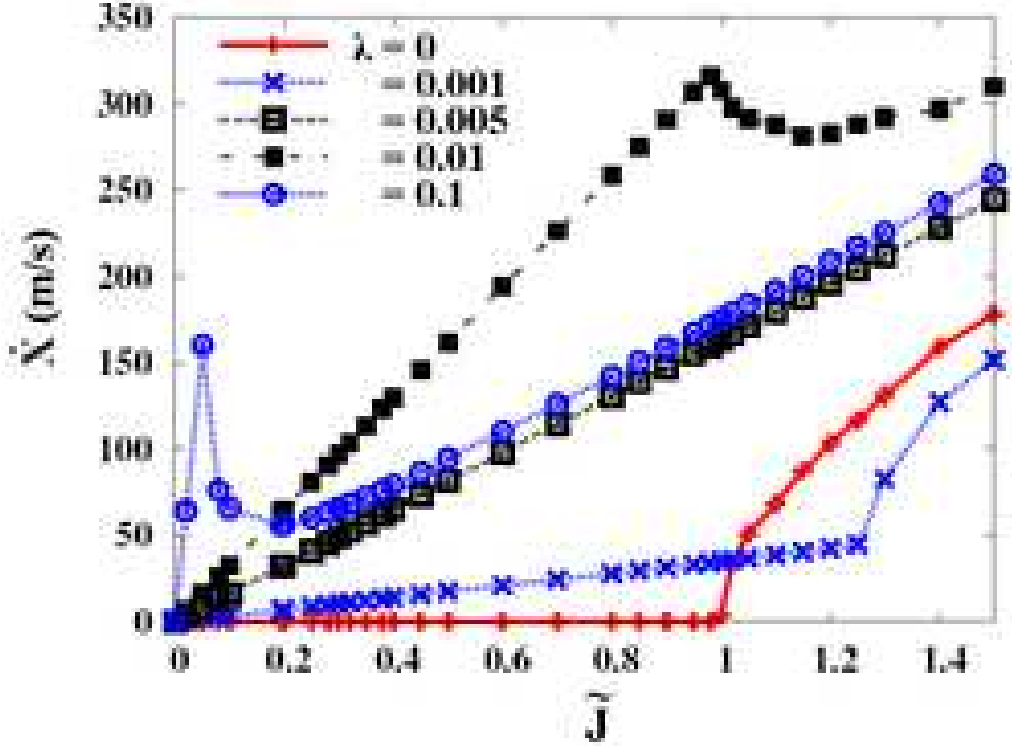}\\
  \caption{Plot of wall velocity as function is current for  $\tla=0,0.001,0.005,0.01$ and $0.1$.
Damping constant is $\al = 0.01$. }
  \label{fig:gravelo}
\end{figure}

We immediately see that Rashba interaction affects Bloch(y) wall  drastically, resulting in vanishing of threshold current and very high velocity. 
This is due to a large effective $\be$ term induced by Rashba interaction, Eq. (\ref{eq:beta67}). 
(Note that we do not consider extrinsic pinning.)
For the present parameters, its ratio to $\alpha$ is given as
$\frac{\beta}{\alpha}=\frac{\tilde{\lambda}}{\alpha}$, and so very large value of $\beta/\alpha\sim 10$ can be realized  
in actual experiment with $\tilde\lambda\sim0.1$ and  $\alpha=0.01$.
Terminal wall velocity of Bloch(y) wall is plotted for different values of $\tla$ in Fig. \ref{fig:gravelo}.
The behavior is consistent with previous studies including $\be$ term~\cite{Thiaville05,TTKSNF06}
indicating that the present motion of Bloch(y) is governed by $\be$ term induced by Rashba interaction.
It is natural since $\be$ term is linear in $\tla$ Eq. (\ref{eq:beta67}) 
while other effects of Rashba interaction enter at the second order (see Eq. (\ref{eq:eyy106})).
Within our analysis, which neglects the extrinsic pinning, threshold current of Bloch(y) is zero due to $\be$ term.
In reality, however, finite threshold will appear from pinning potential~\cite{Thiaville05,TTKSNF06}.
Due to the $\beta$ term, we have obtained very large wall velocity (like 100m/s) at very small current (like 10\% of intrinsic threshold current).  
In reality, Gilbert damping ($\al$) is also modified by Rashba interaction, which might slow the velocity $v$ according to $v\propto\beta/\alpha$\cite{TTKSNF06}.
The modification of $\alpha$ is, however, second order in $\la$~\cite{hanki07,HK08}
 and thus small, hence the high wall velocity would remain unchanged.

\subsection{Bloch(z) and Neel(x) walls}

Behaviors of Bloch(z) and N\'eel(x) walls are essentially the same for present values of parameters, and are distinct from Bloch(y) wall.
These walls have a plateau in $v$-$J$ curve 
for $0.8 \leq \ti{J} \leq 1.3$
as seen in Fig. \ref{fig:grawall01}.
This plateau is due to a step motion of wall induced by the anisotropy field within $\phi$-plane arising from Rashba interaction.
This stepwise motion is induced above threshold current of $\sim 0.8$, and the distance the wall moves is about 22$\mu m$ regardless of current density (for $0.8 \leq \ti{J} \leq 1.3$).

\begin{figure}[htb]
  \centering
  \includegraphics[width=0.3 \linewidth ,clip]{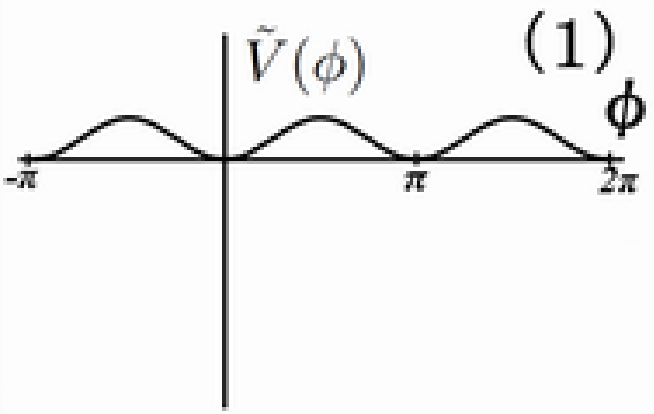}
  \includegraphics[width=0.3 \linewidth ,clip]{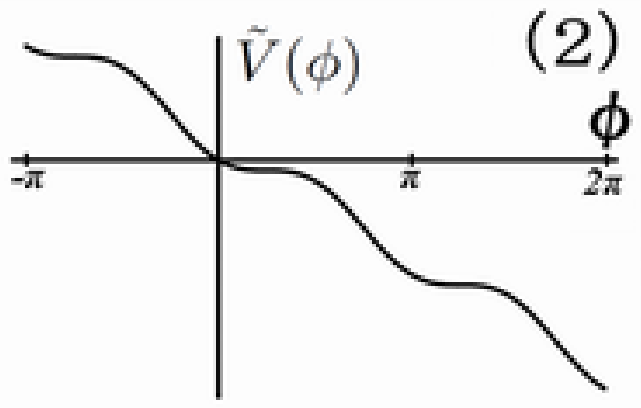}
  \includegraphics[width=0.3 \linewidth ,clip]{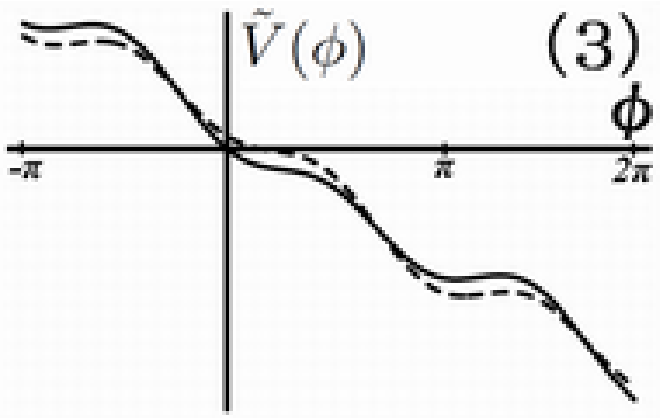}
  \caption{
Potential for $\phi$ of domain wall.
Figure (1) is the potential without current and Rashba interaction, namely potential due to the hard axis anisotropy energy. Energy minimums are at $\phi=0, \pi, \cdots$.
Figure (2) includes the effect of applied current 
below threshold, which tilts the potential 
($\tilde J<0$ here). 
(The intrinsic threshold at $\tilde J=1$ corresponds to the current where local minimum near $\phi\sim0$ disappears.)
The last figure (3) shows the potential in the presence of Rashba interaction, which  lowers (lifts) the energy barrier around $\phi\sim \frac{\pi}{2}$ for 
N\'eel(x) shown by dashed line
(Bloch(z) shown by solid line).
}
  \label{fig:grapote}
\end{figure}

Let us see the mechanism of the plateau in detail.
Dynamics near intrinsic pinning threshold is described by a potential for $\phi$\cite{begt2}, 
obtained from the Lagrangian
(Eq. (\ref{eq:lez274}) and Eq. (\ref{eq:lex275})) as follows:
\eq{
\intertext{Bloch(z):} \ti{V}(\phi) =& \sin^2{\phi} + \frac{\phi}{2 S} \bi{\frac{\Delta}{\Ef} + \frac{\hbar^2}{2 \Delta m \ell^2} \tla^2} \ti{J} 
- \frac{\pi}{2 S} \tla \sin{\phi} \frac{\Delta}{\Ef} \ti{J}. \label{eq:bpz305} \ka
\intertext{N\'eel(x):} \ti{V}(\phi) =& \sin^2{\phi} + \frac{\phi}{2 S} \frac{\Delta}{\Ef} \ti{J} 
- \frac{\pi}{2 S} \tla \cos{\phi} \frac{\Delta}{\Ef} \ti{J}. \label{eq:npx306}
}
These potentials are plotted in Fig. \ref{fig:grapote}.
In the absence of current, the potential is solely by magnetic anisotropy, and energy minimum is at $\phi =0$ 
(Fig. \ref{fig:grapote}(1)).
When current is applied, the potential tilts (Fig. \ref{fig:grapote} (2)).
When local minimum near $\phi = 0$ vanishes, the wall starts to move, 
and this gives the intrinsic threshold current of $\tilde{J_c}=1$ in the absence of Rashba interaction.~\cite{begt2}
When Rashba interaction is switched, $\pi$-periodicity of potential is broken due to the last terms in Eqs. (\ref{eq:bpz305})(\ref{eq:npx306}).
The way of deformation depends on Bloch(z) and N\'eel(x) cases.
Let us consider a  N\'eel(x) wall case where the local energy barrier around $\pi\sim\frac{\pi}{2}$ is lowered by Rashba interaction.
The variable $\phi$ (and wall) starts to move at threshold current (where the local minimum disappears), which is lowered by Rashba interaction.
But this motion stops if current is not large enough since $\phi$ is trapped by the next local minimum near $\phi \sim \pi$,
which has a large energy barrier around 
$\pi\sim \frac{3\pi}{2}$.
Thus in this regime, $\phi$ hops roughly by the amount $\Delta \phi = O(\pi)$ and this corresponds by Eq. (\ref{eq:beta35}) to 
a wall shift of $\Delta X = \frac{\ell}{\al} \Delta \phi$.
This distance corresponds to the distance $\sim 20 \mu$ m seen in Fig. \ref{fig:grawall01}.
If current is sufficiently large to remove the second local minimum,
the wall motion continues and terminal velocity grows 
(for $\tilde{J}>1.3$ in Fig. \ref{fig:grawall01}).
This is the mechanism of plateau. 

An interesting consequence of the step motion is an asymmetric ratchet motion.
We consider a N\'eel wall initially at $\phi=0$ (we choose $\tilde\lambda>0$).
When we apply a negative current, $\tilde J<0$, the local minimum at $\phi=0$ is lifted and the energy barrier for right direction is lowered as we saw in Fig. \ref{fig:grapote}.
In contrast, when current is positive, $\tilde J>0$, the local minimum is lowered ($\propto -\tilde\lambda \tilde J$ by Eq. (\ref{eq:npx306})) and then the effective energy barrier in the  left direction becomes higher.
Therefore, the threshold current for step motion differs by amount $\Delta\tilde J\sim \tilde\lambda$ for the two current directions (starting from fix $\phi$), and thus 
the wall behaves as a ratchet moving only in one way if current is small enough. 
These features are common for Bloch(z) wall case if $\phi=0$ is replaced by $\phi=\pi$.

\begin{figure}[h]
\centering
\includegraphics[width=15cm,clip]{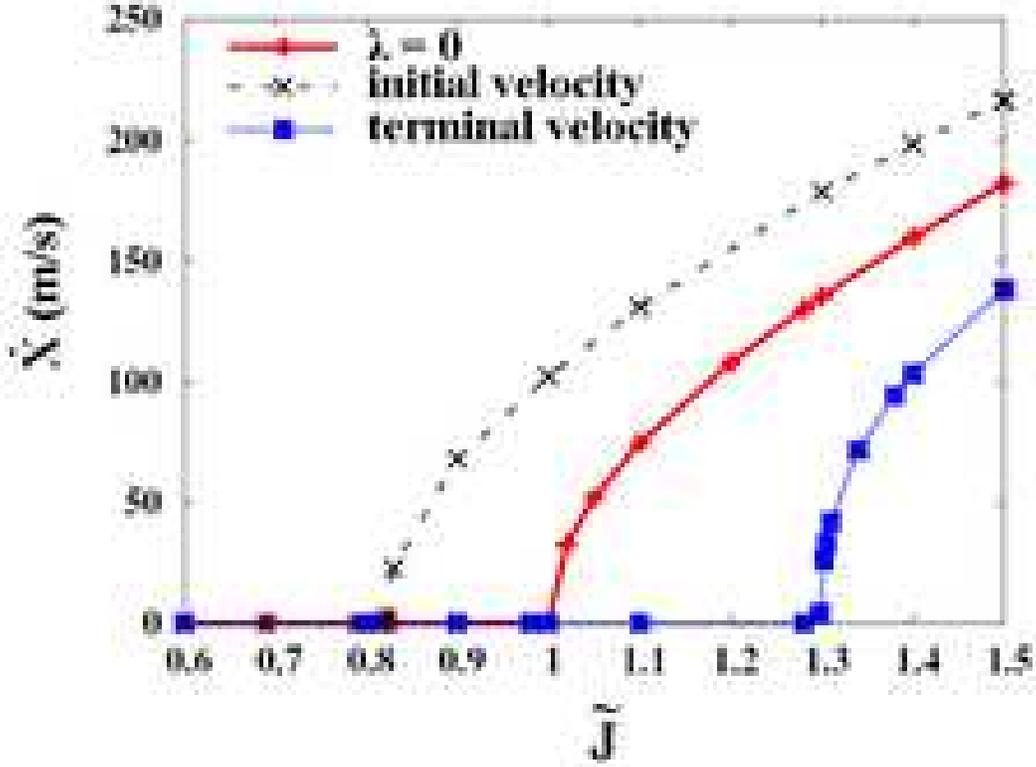}
\caption{The initial and terminal velocity as function of dimensionless current in the case of Bloch(z) and Neel(x) wall.
The velocity is the same for the two walls.
}
\label{fig:bvaz}
\end{figure}

The plateau region has finite initial velocity (but zero terminal velocity).
Figure \ref{fig:bvaz} shows the initial and terminal velocities for Bloch(z) and N\'eel(x) walls
(the velocity is the same for two walls).
We see that even plateau region shows a high velocity comparable to terminal velocity above $\ti{J} = 1$.
For device application, motion over a distance of $20 \mu $m is large enough and so the plateau region would 
be quite useful for low current switching.

\section{Conclusion} 
We have theoretically calculated the effect of the Rashba spin-orbit interaction on the spin transfer torque.
The effective Hamiltonian of local spin under current was calculated using gauge transformation, and the equation of motion for domain wall was derived.
We considered three cases with different magnetic easy and hard axes,
where domain wall structures realized are called Bloch(z), Neel(x) and Bloch(y) walls.
We found there are three influences of Rashba spin-orbit interaction, namely, 
inducing effective magnetic field, increasing spin transfer torque and modification of magnetic anisotropy.
The major effect is that of effective magnetic field,
which arises at the  first-order in Rashba interaction.
Applying voltage in $x$-direction, the effective magnetic field is induced in $y$-direction via Rashba interaction.
In case of Bloch(y) wall, where $y$-direction is the easy axis direction, we showed this field acts as a force which pushes the wall, or in other words, effective $\be$ term arises.
Threshold current thus vanishes. The value of $\beta$ is large even if evaluated for common semiconductor systems, and wall velocity is enhanced greatly.
In contrast, in the cases of Bloch(z) and Neel(x), the effective field is perpendicular the easy axis ($z$ and $x$ directions, respectively),
and step motion of wall over a distance of 
$\Delta X \sim O(\frac{\pi \ell}{\al})$ arises at low current regime,
corresponding to a change of the angle out of easy plane, $\Delta \phi \sim O(\pi)$.
The current necessary for this step motion is lower than the case without Rashba interaction (by $20 \%$ at $\tla = 0.1$).
The initial velocity of step motion is high enough 
(the same order as steady motion slightly above intrinsic threshold).
Wall motion in the step motion regime is asymmetric with respect to current direction, i.e., wall behaves as a ratchet.

Other effects by Rashba interaction, modification of spin transfer torque and anisotropy, appears at second order, $\tla^2$, and are negligibly small.
Change of spin transfer is due to the change of effective electron spin polarization by Rashba interaction.

Rashba interaction arises quite generally when inversion symmetry is broken, e.g., on surface of metals doped with heavy ions\cite{Nakagawa07,Ast07}. 
Such systems would be suitable to realize quite high wall velocity at very small current we have predicted for Bloch(y).

\acknowledgements
The authors are grateful to H. Kohno, J. Shibata, J. Ohe and 
E. Saitoh for valuable discussion.


\end{document}